\documentclass[
    ,final            
  ] {aipproc}

\layoutstyle{8x11single}

\begin{document}

\title{Rheological properties of a granular impurity in the Couette flow}


\author{Francisco Vega Reyes, Vicente Garz\'o and Andr\'es Santos}{
  address={Departamento de F\'isica, Universidad de Extremadura, 06071 Badajoz, Spain}
}

\begin{abstract}
We discuss in this  work the validity of the theoretical solution of
the nonlinear Couette flow for a granular impurity obtained in a
recent work [preprint arXiv:0802.0526], in the range of large
inelasticity and shear rate. We show there is a good agreement
between the theoretical solution and Monte Carlo simulation data,
even under these extreme conditions. We also discuss an extended
theoretical solution that would work for large inelasticities in
ranges of shear rate $a$ not covered by our previous work (i.e.,
below the threshold  value $a_\mathrm{th}$ for which uniform shear
flow may be obtained) and compare also with simulation data.
Preliminary results in the simulations give useful insight in order
to obtain an exact and general solution of the nonlinear Couette
flow (both for $a\ge a_{\mathrm{th}}$ and $a<a_{\mathrm{th}}$).
 \end{abstract}

\maketitle


\paragraph{Introduction}
Granular materials are generically  characterized by the
inelasticity in the collisions between the particles they are
composed of. Transport theories of fluidized granular materials have
been the subject of a considerable amount of research work in the
last few years \cite{aranson06,Goldhirsch}. We may find two
important reasons motivating this effort: 1) the transport of
granular fluids has numerous and obvious industrial applications,
and 2) the theories and experiments on granular media constitute, in
several important cases, generalizations of existing theories on
elastic fluids \cite{aranson06, Goldhirsch}. Additionally,
computational tools used for elastic fluids such as molecular
dynamics (MD) \cite {Rapaport} and Monte Carlo (DSMC) simulations
\cite{Bird} may also be used in the realm of inelastic collisions
\cite{aranson06, Goldhirsch}, and for this reason the study of
granular media is an open and challenging field also for
computational physicists interested in non-equilibrium systems
\cite{aranson06}. We may illustrate the idea of generalization of
previous theories with the help of a paradigmatic example: in the
smooth hard sphere model, inelastic collisions are characterized by
the coefficient of normal restitution $\alpha$, whose value in the
particular case of elastic collisions is unity \cite{Goldhirsch}.
For instance, the recent calculations of transport coefficients of
granular gases for a monocomponent gas \cite{Brey98}, binary dilute
mixtures \cite{VD02}, and the more realistic case of multicomponent
dense mixtures \cite{VDH07} are based on the extension of the
Chapman--Enskog method to inelastic gases, and this is done by
taking into account inelasticity (i.e., the possibility of having
$\alpha\neq 1$) in the collisional integrals of the corresponding
kinetic equation \cite{Brey98}. The multiplicity of behaviors of
granular systems \cite{aranson06} emerges as a very important
consequence of the fact that the study of granular gases is a
generalization of that for elastic gases: for example, while for
elastic gases the steady Couette flow is characterized by a
parameter $\gamma>0$, for granular gases the three possibilities
$\gamma<0, \gamma=0, \gamma>0$ can occur (we will later describe the
meaning of this parameter), which results in having up to 6 types of
steady Couette flows in granular gases whereas in elastic gases only
one is possible \cite{GFT}. Of course this complexity in granular
media arises also in their rheology, whose study is particularly
interesting because nonlinearity is inherent to granular rapid flows
\cite{JStatPhys}, except in the quasielastic limit \cite{GFT}. We
will analyze in this work the rheology of a granular impurity under
Couette flow, focusing on conditions of large inelasticity and
hydrodynamic gradients. We will use for this the exact solution of
the nonlinear Couette flow of a granular impurity, for the case
$\gamma\ge 0$, recently developed by the authors \cite{VGS08.1} and
we will show, with the help of DSMC data, that our theoretical
solution works well even under these extreme conditions.
Nevertheless, our theoretical solution only covers the cases $a\ge
a_\mathrm{th}$, where $a_{\mathrm{th}}$ is the value for which the
well known uniform shear flow may be obtained \cite{JStatPhys},
while the real situation is not restricted to this possibility, as
simulation data we present now clearly show. For this reason, we
will also explore in this work the possibility of extending the
solution to the case $\gamma<0$, which occurs for $a<a_\mathrm{th}$.


\paragraph{Results and discussion}

\begin{figure}
  \includegraphics[height=.225\textheight]{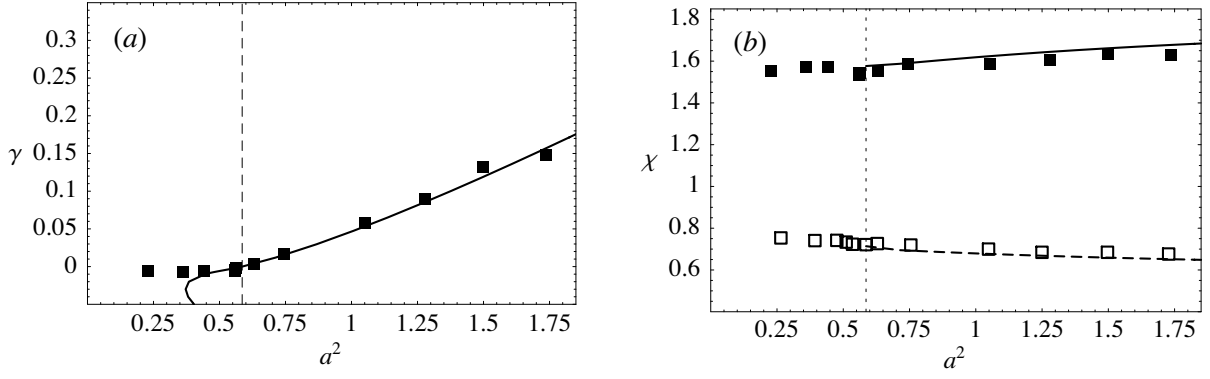}
\caption{(\textit{a}) The curvature parameter $\gamma$ for the
excess component as a function of the reduced shear rate squared
$a^2$. The vertical dashed line indicates the threshold value
$a_\mathrm{th}^2$ separating the regions with $\gamma<0$ (to the
left) and $\gamma>0$ (to the right). The continuous line stands for
the theoretical solution, and the points for DSMC data. The extended
theoretical solution for  $\gamma<0$ works reasonably well in a
finite interval of $a$ to the left of $a_{\textrm{th}}$.
(\textit{b}) The temperature ratio $\chi$ vs. $a^2$ for a granular
impurity with $\alpha=0.5$, a unit size ratio, and mass ratios
$m_1/m_2=2$ (continuous line, solid symbols) and $m_1/m_2=0.5$
(dashed line, open symbols). Lines and symbols stand for theoretical
and DSMC data, respectively. The behavior of $\chi$ clearly
indicates energy non-equipartition even without shear ($a\to 0$).
\label{fig1}}\end{figure}

This work is motivated by a recent previous work \cite{VGS08.1},
where we obtained the exact solution of the nonlinear Couette flow
of  a granular impurity for a BGK-type kinetic model \cite{VS03L},
suitably adapted to the granular binary mixture with the addition of
a drag volume force \cite{VGS07}. We assumed in our previous work
\cite{VGS08.1} that the impurity (species 1) has the same flow
velocity as the excess component (species 2), i.e.,
$\mathbf{u}_1=\mathbf{u}_2$, and that the concentration ratio
$n_1/n_2$ and  the temperature ratio $T_1/T_2$ are spatially
uniform. In addition, the hydrodynamic profiles  of the excess
component \cite{JStatPhys} are an extension of those characterizing
the nonlinear Couette flow for the elastic gas \cite{VS03L}. Thus,
the hydrodynamic profiles have the following form:
\begin{equation}
\frac{1}{\nu_2(y)}\frac{\partial}{\partial
y}u_{2,x}={a}=\mathrm{const}, \quad
\frac{1}{2m_2}\left[\frac{1}{\nu_2(y)}\frac{\partial}{\partial
y}\right]^2T_{2}=-{\gamma}=\mathrm{const},\quad
n_2T_2=\mathrm{const},
\label{hydro}
\end{equation}
where the $x$ axis coincides with the  direction of the flow, the
walls of the system are parallel to it and perpendicular to the $y$
axis,  $\gamma$ is the curvature parameter for the excess component
temperature, and $\nu_2(y)\propto n_2(y)\sqrt{T_2(y)}$ is an
effective collision frequency \cite{VGS08.1}. The solution provides
$\gamma$, the temperature ratio $\chi\equiv T_1/T_2$, and the
rheological properties of both species as functions of the reduced
shear rate $a$ and the mechanical parameters (size ratio, mass
ratio, and coefficients of restitution). In the solution use is made
of the characteristic function $F_{0,m}(\gamma,\zeta_2/\nu_2)$
(where $\zeta_2$ is the cooling rate of species 2) given by
\cite{VGS08.1}
\begin{equation}
F_{0,m}(y,z)= \int_{0}^{\infty} dw\ e^{-(1+z)w} w^m
\left[\sqrt{\pi}\theta(w,y,z)e^{\theta^2(w,y,z)}\mathrm{erfc}\left(\theta(w,y,z)\right)-
1\right],  \quad \theta(w,y,z)\equiv\frac{1}{2\sqrt{2y}} \frac{z}{1-
e^{-\frac{1}{2}z w}}.
\end{equation}
Note that $F_{0,m}(\gamma,\zeta_2/\nu_2)$ is well-defined for
$\gamma\geq 0$ only. As said before, the case $\gamma=0$ corresponds
to a threshold value $a_\mathrm{th}$ of the shear rate at which the
symmetric Couette flow reduces to the uniform shear flow
\cite{VGS08.1,JStatPhys}. On the other hand, one can construct an
\textit{analytical continuation} of the function $F_{0,m}(y,z)$ for
negative $y$ values as
\begin{equation}
\label{0.5} F_{0,m}(y,z)= \int_{0}^{\infty} dw\ e^{-(1+z)w} w^m
\left[\sqrt{\pi}\widetilde{\theta}(w,y,z)e^{-\widetilde{\theta}^2(w,y,z)}\mathrm{erfi}\left(\widetilde{\theta}(w,y,z)\right)-
1\right], \quad \widetilde{\theta}(w,y,z)\equiv\frac{1}{2\sqrt{-2y}}
\frac{z}{1- e^{-\frac{1}{2}z w}}.
\end{equation}

\begin{figure}
  \includegraphics[height= .222\textheight ]{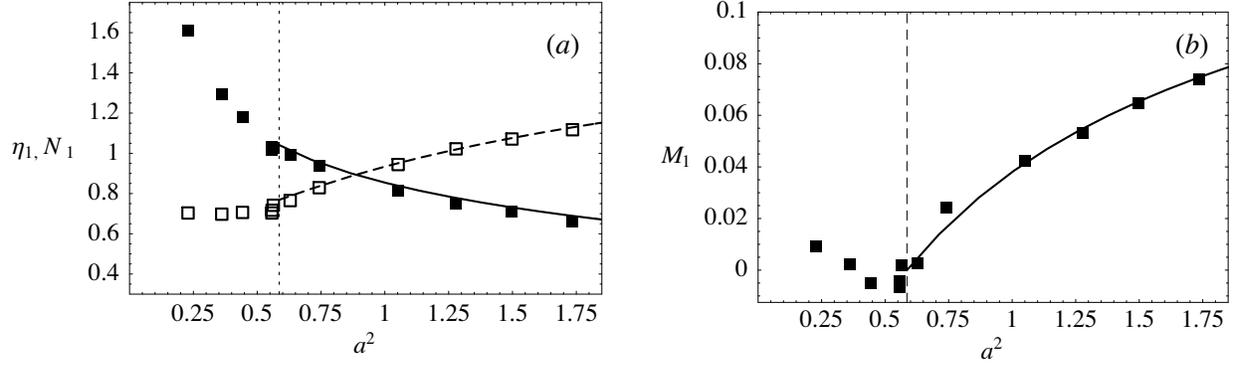}
\caption{(\textit{a})  Viscosity $\eta_1$  (continuous line, solid
symbols) and normal stress difference $N_1$ (dashed line, open
symbols) coefficients vs. $a^2$ for a granular impurity with
$\alpha=0.5$, a unit size ratio, and a mass ratio $m_1/m_2=2$.
(\textit{b}) Normal stress difference coefficient $M_1$ vs. $a^2$
for the same granular impurity. Lines stand for theory and points
for simulation data.\label{fig2}}
\end{figure}

We proved the hypotheses $\mathbf{u}_1=\mathbf{u}_2$,
$n_1/n_2=\mathrm{const}$, $T_1/T_2=\mathrm{const}$, and Eq.\
\eqref{hydro}  to be accurate by numerically solving (DSMC) our
model kinetic equations, at least for values of the coefficient of
normal restitution as low as $\alpha=0.8$ and values of the reduced
shear rate as high as $a\sim 1$ \cite{VGS08.1}. We present in this
work results for considerably larger inelasticities ($\alpha=0.5$)
and shear rates (up to $a=1.75$). Additionally, we present
preliminary results, for the excess component, of an approximate
solution  based on Eq.\ \eqref{0.5} of the nonlinear Couette flow in
the $\gamma<0$ region that we are currently working out. As we see
in Figs.\ \ref{fig1} and \ref{fig2}, the results are quite
satisfactory: in Fig.\ \ref{fig1}(\textit{a}) we can notice that the
extended solution works well for $\gamma(a)$ in a finite range of
values of $a<a_{\mathrm{th}}$, although the solution evidently fails
below a critical value of $a$, and that the regular solution for
$a\ge a_{\mathrm{th}}$ ($\gamma\ge 0$)  shows an excellent agreement
with DSMC data. It is also to be noticed the good agreement for the
temperature ratio $\chi=T_1/T_2$, that, as it is known and we can
see in the figure, in granular mixtures is different from 1 even in
the absence of shear ($a=0$). In Fig.\ \ref{fig2} we present the
results corresponding to the stress tensor transport coefficients of
the impurity. All of them show a very good agreement with the
numerical solution: the non-Newtonian shear viscosity $\eta_1$, and
the normal stress differences $N_1=(P_{1,xx}-P_{1,yy})/p_1$,
$M_1=(P_{1,zz}-P_{1,yy})/p_1$, with $p_1=n_1T_1$, whose magnitude
shows the importance of rheological effects in this system. We can
also notice in Figs.\ \ref{fig1} and \ref{fig2} that, independently
of the limitation of our theory to the case $a\ge a_\mathrm{th}$,
the numerical solution of the kinetic equation (simulation data)
extends beyond that limit, and the functions represented do not show
any discontinuity in their behavior (except perhaps $M_1$) at
$a=a_\mathrm{th}$.

Summarizing, we have shown in this work that the  theoretical
solution of the nonlinear Couette flow obtained in a previous work
\cite{VGS08.1} can be safely used for large inelasticities
($\alpha=0.5$) and hydrodynamic gradients ($a\sim 1.75$). We also
checked that the hypotheses underlying the theoretical solution, as
well as its predictions for the rheological properties, are
fulfilled for this large inelasticity.  We have also discussed an
extension of this theoretical solution in the region of shear rates
below the threshold value $a<a_{\mathrm{th}}$, i.e., negative value
of the curvature parameter $\gamma$. These results are also useful
for a future work, where we will assess to which degree our BGK-type
kinetic model is able to describe the results from the 'true'
Boltzmann kinetic equations.

\begin{theacknowledgments}
This research has been supported by the Ministerio de Educaci\'on y
Ciencia  (Spain) through Programa Juan de la Cierva (F.V.R.) and
Grant No.\ FIS2007-60977, partially financed by FEDER funds, and by
the Junta de Extremadura (Spain) through Grant No.\ GRU08069.
\end{theacknowledgments}

\bibliographystyle{aipproc}   

\bibliography{VGSicr08}
\end{document}